# Magnetic-field enhancement of performance of superconducting nanowire single-photon detector

Ilya Charaev, Alexei Semenov, Konstantin Ilin, Michael Siegel

*Abstract*— We present SNSPDs from NbN nanowires shaped after square-spiral that allows an increase not only in critical currents but also an extension of spectral detection efficiencies by just applying an external magnetic field. Using negative electron-beam lithography with the positive resist for shaping nanowires, made it possible to reduce the inner bend radius. Consequently, the effect of critical-current enhancement in the magnetic field becomes stronger than it was demonstrated earlier. Here we achieved a 13% increase of the critical current in the magnetic field. We measured spectra of the single-photon detection efficiency in the wavelength range from 400 to 1100 nm in the magnetic field. At zero field, the square spiral has the spectrum similar to that of a meander. At the field providing the maximum of the critical current, the detection efficiency and the cut-off wavelength in the spectrum increase by 20 % and by 54%, correspondingly. The magnetic-field dependence of dark count rate is well described by proposed analytical model.

## I. INTRODUCTION

**M**AGNETIC vortices with their fluctuations, dynamics and interaction with each other are essential part of different theoretical models explaining experimental observations and detection mechanism of superconducting nanowire single-photon detectors (SNSPDs) [1]. Vortices and antivortices are present even at zero magnetic field, particularly in dirty, strongly type-II superconductors [2]. The single-vortex crossing is considered as a possible mechanism of photon and dark counts in SNSPDs. At wavelength beyond the cut-off wavelength $\lambda_C$ [3], the energy of absorbed photon is not sufficiently large to create normal state in the nanowire. However, thermodynamic fluctuation may further suppress superconductivity in the nonequilibrium spot and initiate nucleation of a vortex. The Lorentz force imposed by the current drives the vortex across the nanowire over the potential barrier. When moving, vortex dissipates energy along its trajectory and creates a resistive belt which gives rise to the normal domain (vortex-assisted photon count event). A much stronger fluctuation is required to initiate vortex crossing in the absence of photons that produces a dark count. Dark counts are therefore rare and appear only at bias currents close to the critical current.

Theoretical study of current crowning in bends of superconducting nanowires revealed field-induced increase of critical current in bends of superconducting nanocircuits [4] that was experimentally confirmed [5]. Several experimental observations have been reported on photon and dark count rates, *PCR* and *DCR*, for SNSPD in magnetic field. For meander SNSPD, a slight asymmetry in *DCR* was found with respect to the field-direction [6]. It was assigned to the difference in shapes of adjacent turns of the meander. Renema *et al.* [7] found enhancement of the intrinsic detection efficiency in a straight nanowire in magnetic field. Lusche *et al.* [8] demonstrated an enhancement of *PCR* and *DCR* for meander-shaped SNSPDs in external magnetic field. In this experiment, the ratio of the bias current $I_b$ to the critical current in magnetic field $I_C(B)$ increased with the increase of magnetic fields. In a similar experiment, reduction of *PCR* in magnetic field was found for meanders at fixed relative bias current $I_b/I_C(B)$ [9]. In all experiments mentioned above, the measured critical current of either meanders or straight strips decreased with the increase of the magnetic field. Because of the opposite symmetry of adjacent turns in the meander with respect to the directions of the field and the flow of the bias current, uniform external magnetic field enhances current crowding effect in one turn and reduces it in the adjacent turn. Hence, for any field direction, the energy barrier for vortex penetration decreases in one half of turns. The same occurs in straight strips. To overcome the reduction of the critical current at small fields, the detector layout (square spiral) with the same symmetry for all bends was proposed [10], [11]. It has been shown for square-spiral SNSPDs that magnetic field increases the critical current and *PCR* and decreases *DCR* to a minimum of less than <10 cps at the field which corresponds to the maximum of $I_C$ [11]. Although the performance enhancement in magnetic field was successfully demonstrated for such detectors, some questions remain unanswered: (1) What is the relation between $\lambda_C$ and the enhanced critical current in magnetic field?; (2) Does the magnetic field affect the detection efficiency with an increase of the bias current? Intuitively, the evolution of the spectra of detection efficiency with the increase of the magnetic field is expected. Different models of SNSPD response predict dependences of $\lambda_C$ on the ratio of bias current to the de-pairing current $I_C^d$. In spite of different details of the predictions, all models agree that the increasing ratio should cause a decrease of the minimal detectable energy of photon.

Ilya Charaev is with the Department of Electrical Engineering and Computer Science, Massachusetts Institute of Technology, Cambridge, MA 02139, USA (e-mail: charaev@mit.edu).
Alexei Semenov is with the Institute of Optical Systems, German Aerospace Center (DLR), Berlin, 12489, Germany (e-mail: Alexei.Semenov@dlr.de).
Konstantin Ilin, Michael Siegel are with the Institute of Micro- und Nanoelectronic Systems, Karlsruhe Institute of Technology (KIT), Karlsruhe, 76187, Germany (e-mail: konstantin.ilin@kit.edu, michael.siegel@kit.edu).

In this paper, we demonstrate the method to enhance the detection efficiency and extend the cut-off wavelength towards the infrared range. We also show that the usage of the unconventional negative electron-beam lithography with the positive resist makes the effect of magnetic field stronger.

## II. TECHNOLOGY

### A. *Thin-film deposition and characterization*

A niobium nitride thin film was deposited on an R-plane cut, one-side polished, squared 10 x 10 mm² sapphire substrate by reactive magnetron sputtering. The film was deposited from the 2-inch pure Nb target (99.95%) in an argon (Ar) and nitrogen ($N_2$) atmosphere on wafers which were placed without any thermal glue on a heater at 850°C. The partial pressure of argon and nitrogen was $P_{Ar} = 1.9 \times 10^{-3}$ mbar and $P_{N2} = 2.7 \times 10^{-4}$ mbar, respectively. At the discharge current 150 mA the deposition rate of NbN was 0.07 nm/s. These conditions ensure the particular stoichiometry of NbN which results in the highest transition temperature. The film had a thickness of $d \approx$ 5 nm, which was measured by a stylus profilometer.

The temperature dependence of the film resistance was measured immediately after deposition in the range from 300 down to 4.2 K by the standard four-probe technique. The square resistance of the film was 255 Ω/square at 30 K. The critical temperature $T_c$ was defined as the lowest temperature at which a non-zero resistance could be measured. The measured critical temperature of the film was about 14 K (Fig. 1). The residual-resistance ratio (*RRR*) was evaluated as the ratio between the resistance at room temperature to the maximum resistance at $T = 20$ K: $RRR = R_{300}/R_{max} = 0.95$. The resistivity, ρ, was evaluated with the measured thickness, $d$, and the square resistance, $R_s$, of the films at $T = 20$ K as $\rho = d\, R_s$. For our film we found the resistivity 122 μΩ×cm. The temperature dependence of the second critical magnetic field $B_{C2}(T)$ was measured by applying an external magnetic field ranging from zero to 1.2 T perpendicularly to the film surface (insert in Fig. 1). The critical field was associated with the field driving the film resistance to the half of the normal state value. Using the theoretical dependence of $B_{C2}(T)$ in the dirty limit, we estimated the value of the coherence length at zero temperature as

$$\xi(0) = \sqrt{\frac{\Phi_0}{2\pi B_{C2}}} \qquad (1)$$

that resulted in $\xi(0) = 4.1$ nm for our NbN film. Electron diffusion coefficient was computed from the slope of the linear part of the dependence $B_{C2}(T)$ near $T_C$ with the formula

$$D = -\frac{4k_B}{\pi e}\left(\frac{dB_{C2}}{dT}\right)^{-1} \qquad (2)$$

that gave the electron diffusion coefficient $D = 0.54$ cm²/s.

With the experimental value of superconducting energy gap in NbN $\Delta(0) = 2.05\, k_B\, T_C$, we found the magnetic penetration depth

$$\lambda(0) = \sqrt{\frac{\hbar \rho}{\pi \mu_0 \Delta(0)}} \;. \qquad (3)$$

and the Pearl length $\Lambda = 2\lambda(0)^2/d$ to amount to $\lambda(0) = 289$ nm and $\Lambda = 34$ μm for our NbN film.

### B. *Detector fabrication*

The film was patterned into the square spiral *via* negative electron-beam lithography over the PMMA resist [12] and, subsequently, *via* ion beam milling at an argon pressure of $1.2\times 10^{-4}$ mbar. The variation of the dose and the width of nanowires was done to find the limit on the resolution in turns and to maximize the effect of critical-current enhancement by the magnetic field in square-spiral SNSPD in accordance to theoretical considerations proposed in Ref. [4]. The nanowires were exposed with dose 10000 μC/cm². The exposed structures were developed in acetone for 1.5 min and then rinsed in 2-propanol while the standard developer for 45 s (30% MIBK in 2-propanol at 22°C) was used for developing of positive pattern. The geometry was controlled by SEM inspection. The Fig. 2 represents SEM-example (left image) of square spiral nanowires. The width of the nanowire was 89 nm. We measured the inner radius of turns in the square spiral to be 35 nm.

To lead bias current through, the spiral was isolated from the top except for the central pad and then a top electrode was brought above the isolating layer. The sketch (Fig. 2, right panel) shows a design of multilayer square-spiral SNSPD. All dimensions of the device and technological aspects of fabrication have been reported earlier [10].

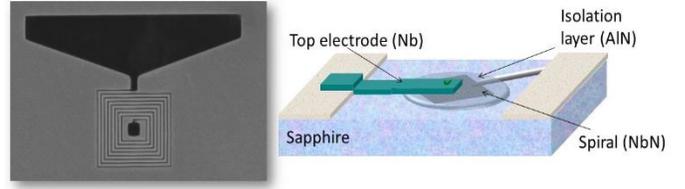

Fig. 2. SEM image of square spiral after development process (left). Schematics of the multilayer structure with the isolator and the top electrode (right).

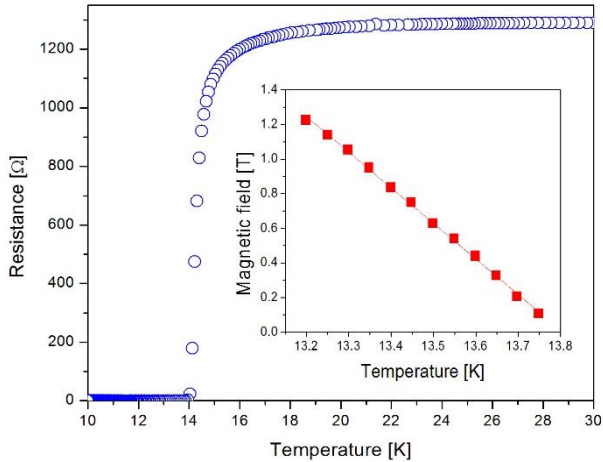

Fig. 1. *R(T)*-dependence of 5-nm thick NbN film on sapphire substrate. Insert: $B_{C2}(T)$-dependence of the film.

## III. MEASUREMENTS

### A. Experimental setup

The measurement setup includes a dipstick with a superconducting solenoid and several readout components for operation at room temperature. The dipstick with the magnet was immersed in a Dewar with liquid helium; it includes a vacuum chamber and a sample rod on which the SNSPD is installed. To measure the actual field near the sample, a calibrated Hall sensor with ultra-high sensitivity and wide measuring range (1 mT - 2T) was mounted on the sample holder. The low-temperature bias tee decouples the high frequency path from the DC bias path. The high-frequency signal is led out of the dipstick by stainless-steel rigid coaxial cables, while DC bias is provided via a pair of twisted wires. The sample is biased by a battery-powered low-noise DC source. The signal is amplified at room temperature by several amplifiers with the total gain of 70 dB and then led to a pulse counter with a 300 MHz physical bandwidth. The optical fiber feeds the light from the monochromator into the cryogenic part and is mounted on a movable stage above the sample surface. Detector is illuminated from the rear side, i.e., through the non-polished sapphire substrate.

### B. Results

The measurements of the critical current in the magnetic field were performed at 4.2 K. The $I_C(B)$-behavior is asymmetric with respect to zero magnetic field. Fig. 3 represents results obtained for square spiral SNSPD. The critical current of the detector with a wire width of 90 nm was 48 µA at $B = 0$ T. The measured critical current of the structure is reduced by current crowding [13] at the inner corners of bends with respect to the critical current of the straight parts. The external magnetic field induces screening current in bends. By applying small magnetic field, the critical current was continuously increased and reached the maximum of 54.1 µA at $B_{max}=56$ mT. At $B > B_{max}$, the supersession of the current critical was observed. The maximum of the critical current in magnetic field was found to be ~13% higher than $I_C(0)$.

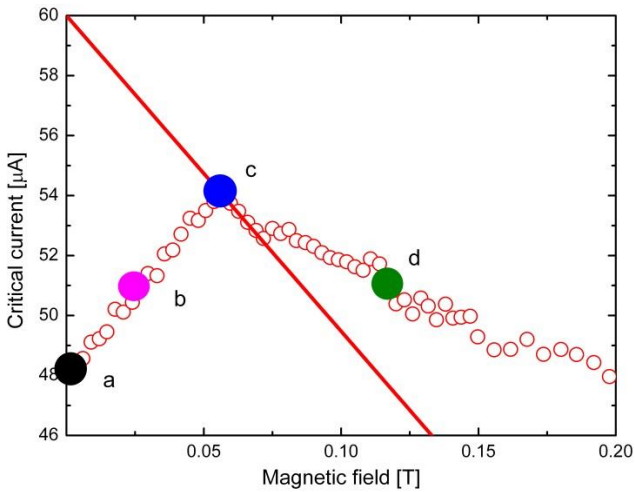

Fig. 3. Critical current of square-spiral SNSPD vs magnetic field. Colored circles indicate fields at which spectral measurements and DCR measurements were performed. Solid line represents the dependence $I_C(B) = I_{C0} (1- B/ B_S)$ with $I_{C0} = 60$ mA and $B_S = 570$ mT.

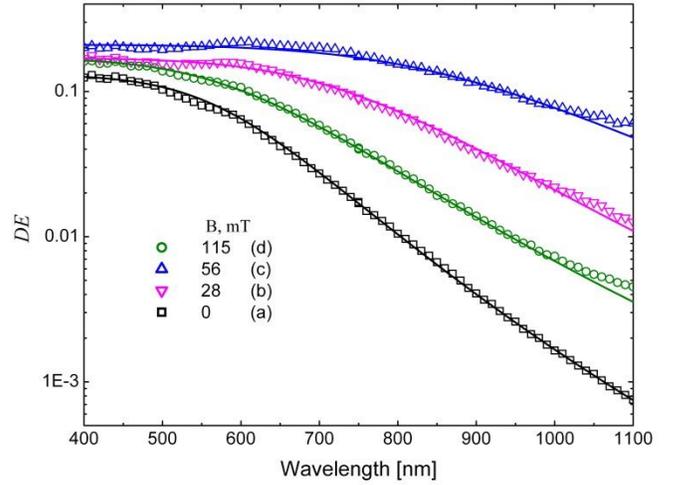

Fig. 4. Spectral detection efficiency of square-spiral SNSPD at different magnetic fields indicated in the legend. Measurements were performed at relative bias current $I_b = 0.95\ I_C(B)$ at 4.2 K. Solid lines show fits with Eq. 4.

The spectra of the detection efficiency (*DE*) of square-spiral SNSPDs have been measured in the wavelength range from 400 up to 1100 nm at different magnetic fields. We chose four field values shown with colored circles in Fig. 3: (a) zero magnetic field; (b) 28 mT; (c) $B_{max}=56$ mT; (d) 115 mT. The results were obtained at the fixed relative bias current $I_b = 0.95\ I_C(B)$. For the particular wavelength, the *DE* was defined as the ratio of the rate of photon counts to the rate of photon arrival to the illuminated side of the detector. The transition from the plateau to the decaying part of the *DE* spectrum beyond the cut-off (Fig. 4) is formally described by:

$$DE(\lambda) = \left(1 + \left(\frac{\lambda}{\lambda_C}\right)^p\right)^{-1} \quad (4)$$

where *p* describes the power-law decrease of the efficiency in the near infrared range.

At $B = 0$ T, the cut-off wavelength $\lambda_C = 600$ nm is very close to the value obtained for meander-type SNSPD. At the maximum $I_C$ in the magnetic field, the detection efficiency at the plateau grew up by 20 %. The DE spectrum broadened to $\lambda_C = 925$ nm which corresponds to a 54% increase of the cut-off wavelength. Although the critical current of the detector is the same at $B = 28$ mT and $B = 115$ mT, the cut-off wavelength was 770 and 635 nm at the smaller and at the larger magnetic field, respectively.

The dark-count rate was determined by blocking the optical path at the warm edge of the fiber. *DCR* was investigated in magnetic field at bias currents from 0.9 up to $0.99I_C(B)$. Fig. 5 represents results of measurements for square-spiral detector. The minimum of *DCR* was found at 56 mT to be ~$10^2$ cps which corresponds to the maximum critical current in the magnetic field. At $B = 115$ mT, the rate of the dark count is ~$10^3$ cps which is similar to *DCR* at zero field.

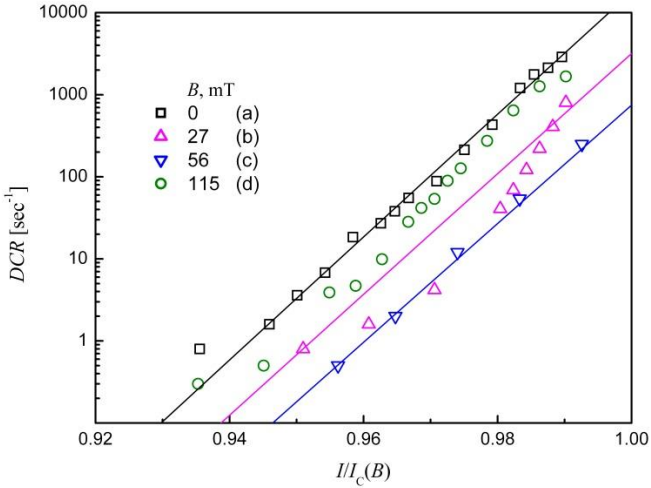

Fig. 5 The dark count rate versus relative bias current presented for different magnetic fields indicated in the legend. Solid lines present simulations in the framework of the vortex model described in the text.

*C. Discussion*

Qualitative interpretation of our experimental data is quite straightforward. Magnetic field of the appropriate direction with respect to the current flow reduces current crowding in all bends in the square spiral. This allows one to apply a larger bias current in the superconducting state that leads to an increase of the spectral bandwidth [14]. However, the field simultaneously decreases the critical current in straight portions of the spiral. The maximum in the critical current is achieved when the decreasing critical current in straights meets the increasing critical current in bends. Extrapolation of the linear decrease in $I_C(B)$ after maximum to $B = 0$ gives the critical current of straights at the zero field $I_{C0}$ [4].

It has been argued that bends and turns with sharp corners are places from which dark counts mostly originate [10]. In our case, magnetic field reduces the local current density at the inner sides of bends and flattens the current density profile across bends. The effective barrier for vortices or phase slips increases and that decreases the rate of dark counts coming from bends. *DCR* decreases with the field until the critical current in straights drops below the critical current in bends. Since the total length of straights is much larger than the total length of bends, the former dominate in delivering dark counts at large fields. *DCR* further increases as the field suppresses the energy barrier in straights.

Here we propose a simple analytical model of vortex hopping in bends which quantitatively describes our experimental data. We neglect the asymmetry in the current circulating the vortex in the bend but retain the effect of crowding for the bias current and the current induced by external magnetic field. The profile of the current sheet density $K$ in the bend is obtained from the joint solution of Maxwell and London equations resulting in $\nabla \cdot K = 0$ and $\nabla \times K = -2B/(\mu_0 \Lambda)$. We assumed that around the apex of the bend the radial component of $K$ equals zero and that the tangential component depends only on the distance $r$ from the geometric center of the bend. For the applied bias current $I$ and the field $B$ the sheet density of the total current in the bend with radii $a$ and $b$ and the width $w = b - a$ is

$$K(r) = r^{-1} \ln\left(\frac{b}{a}\right)^{-1} \left(I + \frac{Bw(w+2a)}{\mu_0 \Lambda}\right) - \frac{2B}{\mu_0 \Lambda} r \ . \quad (5)$$

The height of the barrier for vortices is then obtained by integrating the total force on the vortex $F_0 + K(r) \Phi_0$ where $\Phi_0$ is the magnetic flux quantum and $F_0 = -\Phi_0^2/(2\mu_0 \Lambda w)$ ctg$(\pi r/w)$ is the force on the vortex imposed by the boundary conditions. The critical values of the current and fields are defined as values suppressing the barrier. To agree the model predictions with the experimental $I_C(B)$ dependence we had to cut the barrier at a distance $0.7\xi$ from the bend edges and to adopt an effective superconducting width of the strip of 67 nm. This results in the critical current of straits $I_{C0} = 60$ μA and in the critical field $B_S = 570$ mT. The dependence $I_C(B) = I_{C0} (1 - B/B_S)$ is shown with the solid line in Fig. 3. We computed *DCR* using the formalism (Eq. 4 in [15]) and the model value of the barrier. The model predictions for *DCR* at different fields are shown in Fig. 5 with solid lines. With the given scattering of experimental points the agreement between the experiment and the model is very good.

IV. CONCLUSION

We demonstrated a method to enhance the detection efficiency and spectral cut-off of superconducting nanowire single-photon detectors by external magnetic field. Using one-symmetry square-spiral geometry and unconventional negative-PMMA lithography, we achieved a 13% increase of the critical current for the device with the nominal width of 90 nm. Magnetic field enhanced the detection efficiency by 20% and increased the wavelength corresponding to the edge of the spectral band from 600 nm to 1 μm. We proposed a simple quantitative model relying on the barrier for vortex hopping in bends which nicely describes variation of the critical current and the dark count rate in magnetic field. The agreement further supports earlier findings that bends are dominating source of dark counts in bended nanowires.